\documentclass[twocolumn,english,aps,pra,superscriptaddress,amsmath,amssymb,floatfix,nofootinbib,longbibliography
]{revtex4-2}

\usepackage{amsthm}
\usepackage{amsfonts}
\usepackage{siunitx}
\usepackage{amsmath}
\usepackage{amssymb}
\usepackage{graphicx}
\usepackage{verbatim}
\usepackage[colorlinks]{hyperref}
\usepackage{tikz}
\usepackage{pgfplots}
\usepackage{adjustbox}
\usepackage{braket}
\usepackage{xcolor}
\usepackage{physics}
\usepackage{amssymb} 
\usepackage{graphicx}
\usepackage{dcolumn}
\usepackage{bm}
\usepackage{mathtools}
\usepackage{hyperref}
\usepackage{mathrsfs}
\usepackage{dashrule}
\usepackage{caption}
\usepackage{subcaption}
\usepackage{quantikz}
\usepackage[version=4,arrows=pgf-filled,
textfontname=sffamily,
mathfontname=mathsf]{mhchem}

\usepackage[font=small,labelfont=bf,
   justification=justified,
   format=plain]{caption}

\definecolor{linkcolor}{RGB}{0,83,166}
\hypersetup{
  colorlinks = true,
  allcolors = {linkcolor}
}

\begin{document}

\title{VQE as Initial State Preparation for QPE on Heisenberg Spin-Glass Hamiltonians}

\author{Elijah Pelofske}
\email[]{epelofske@lanl.gov}
\affiliation{Information Systems \& Modeling, Los Alamos National Laboratory}
\affiliation{Quantum \& Condensed Matter Physics, Los Alamos National Laboratory}
\affiliation{Center for Quantum Computing, Los Alamos National Laboratory}

\author{Stephan Eidenbenz}
\email[]{eidenben@lanl.gov}
\affiliation{Information Sciences, Los Alamos National Laboratory}
\affiliation{Center for Quantum Computing, Los Alamos National Laboratory}

\begin{abstract}

Quantum Phase Estimation (QPE) is the quantum algorithmic workhorse for computing ground state energies of quantum Hamiltonians with quantum computers. Ground state energy calculation of physical systems is perhaps the most promising use case for quantum computing in terms of scientific and commercial value with a plausible path to outperformance of classical alternatives. This path, however, hinges on the availability of initial states for QPE with significant overlap with the true ground state. Using extensive (classical) numerical computations, we study whether the NISQ-era algorithm VQE (Variational Quantum Eigensolver) could be used to efficiently prepare high-overlap states of disordered fully-connected anisotropic Heisenberg spin glass quantum Hamiltonians with up to $15$ qubits. 
We find that (i) -- consistent with widely held, but rarely numerically illustrated  beliefs -- VQE is generally unable to efficiently converge to the ground state for our Hamiltonians, which is a well-known issue with VQE due to a variety of factors including vanishing gradients and local minima; (ii) low energy states do not necessarily have large ground-state overlap, but there is typically a correlation between the two measures; (iii) adding more than three layers to the VQE ansatz neither improves overlap nor the energies found; and (iv) the best-found overlap scaling as a function of the Hamiltonian system size is not strongly exponentially decreasing, suggesting potential for VQE to be a heuristic state preparation algorithm for QPE.

\end{abstract}

\maketitle

\section{Introduction}
\label{section:Introduction}

Quantum Phase Estimation (QPE)~\cite{kitaev1995quantummeasurementsabelianstabilizer, nielsen2010quantum} is a foundational quantum algorithm that features significant speedups for many fundamental calculations. QPE is not only a subroutine in several key quantum algorithms including HHL~\cite{PhysRevLett.103.150502} and Shor's factoring algorithm~\cite{365700, Shor_1997}, but is also by itself an efficient algorithm for computing the minimum eigenvalue of quantum Hamiltonians. This core computation is crucial for many fields of science including electronic structure problems. Executing QPE as a digital quantum algorithm requires error corrected quantum computers that are significantly larger than current ``NISQ'' computers~\cite{Preskill_2018}. Algorithmically, the primary open question for the practical usage of QPE is initial state preparation\footnote{Initial state preparation with high ground-state overlap is also important for several other quantum algorithms besides QPE~\cite{yu2025quantumcentricalgorithmsamplebasedkrylov}, but QPE is arguably the most consequential}. Standard QPE accuracy, and moreover the digitized minimum eigenvalue probability, depends only on three factors (for a fixed number of bits of precision): the time parameter $t$ in the time evolution $e^{iHt}$, the accuracy with which an approximate evolution of $e^{iHt}$ can be implemented, and the overlap between the initial state and the ground state(s) of the target quantum Hamiltonian. Here, we consider the third factor: initial state overlap. Initial state preparation that has a high overlap with a quantum Hamiltonian ground state, which itself may be degenerate, is computationally very challenging in general. Since QPE's probability of measuring the eigenstate is proportional to the initial-state overlap and the total evolution time~\cite{pelofske2026numericalexperimentsparametersetting, kaye2006introduction}, an exponentially small (in number of qubits) overlap will result in an exponential number of QPE runs required to get the ground state with a constant probability; thus one would ideally want to guarantee an overlap of constant or at least only polynomially small size in order to reap QPE's speedup benefit.

Initial state preparation can suffer from the orthogonality catastrophe~\cite{PhysRevLett.18.1049}, which can lead to exponentially decreasing overlaps~\cite{Louvet_2026, Lee_2023}. For this reason, high-quality state preparation for quantum Hamiltonians is a subject of considerable interest for the quantum algorithms and quantum computing communities. Generally, this is a very hard problem to address as of course any guarantees of polynomially small overlaps achieved by such a state preparation algorithm could be used to construct a quantum algorithm that, when combined with QPE, would be guaranteed to find ground state energies, which is a $QMA$-complete problem for even 2-local Hamiltonians \cite{kempe2006complexity}. Recall that $QMA$ is the quantum analogue of the classical $NP$ complexity class. 

Setting aside worst-case complexity bounds, it is an open debate as to how hard or easy it is to prepare initial states with high ground state overlaps for practical application domains with the general assumption being that the practical complexity of finding high overlap states increases with the correlation or entanglement levels of the ground states. However, there are very few numerical studies on this question. In this study, we examine how suitable the Variational Quantum Eigensolver (VQE) is for finding high overlap ground states for the case of fully-connected Heisenberg Hamiltonians. While a more thorough study would extend to other classes of Hamiltonians and state preparation methods (such as adiabatic state preparation, or domain specific approaches such as Hartree-Fock or Coupled-Cluster states for electronic structure Hamiltonians), we uncover interesting structural facts in the Heisenberg model case. 

Using energy-based cost function models is in general a very powerful tool for exploring computationally challenging search spaces. The Variational Quantum Eigensolver (VQE)~\cite{Peruzzo_2014} is one such energy based method. VQE is a now well-established hybrid quantum algorithm that -- just like QPE -- aims to compute minimum eigenvalues of exponentially large matrices (e.g., quantum Hamiltonians) allowing a wide range of quantum simulations to be performed~\cite{PhysRevLett.122.230401, PhysRevB.108.075127, Kandala_2017, Stanisic_2022, bosse2021probinggroundstateproperties, perez2023nuclear, Angelides_2025, carolan2026sigmavqeexcitedstatepreparationquantum, Uvarov_2020, weaving2025simulatingantiferromagneticheisenbergmodel, kirmani2025variationalquantumsimulationstwodimensional}. VQE is considered a ``NISQ'' friendly in-between which is similar in its goal to QPE, but can be run using very short depth quantum circuits and therefore is practical to implement on current quantum computing hardware. The tradeoff is that parameters must be learned using classical machine learning and optimization algorithms, in an iterative learning loop with the quantum hardware. However, several key obstacles have been identified in VQE training, besides the standard combinatorial complexity of searching a high-dimensional parameter space. These obstacles include a combination of vanishing gradients, local minima proliferation, qubit decoherence, and shot noise that make classical learning very challenging~\cite{Arrasmith_2021, Ragone_2024, Cerezo_2021, Anschuetz_2022, Wang_2021, McClean_2018, Larocca_2025, Uvarov_2021}. Therefore, the scalability of VQE, to larger quantum Hamiltonians while maintaining high-accuracy for quantum Hamiltonian ground-state calculation, is unclear. 

Moreover, using the energy based cost function also has a fundamental limitation for state overlap in that low-energy is a \emph{necessary but not sufficient condition}. There is no guarantee that a state generated by VQE will have a high overlap, but at the same time, energy based cost functions are effectively the only method currently available for approximately preparing states with high overlap~\cite{Louvet_2026, Fomichev_2024}. 

The training of variational quantum algorithms to fully obtain the true ground-state is generally quite hard to achieve, even in ideal settings, instead algorithms like VQE can typically only approximately prepare ground-states~\cite{O_Malley_2016, Mih_likov__2022, kirmani2025variationalquantumsimulationstwodimensional, Lively_2024, belaloui2024groundstateenergyestimation, tripathi2026ansatzexpressivityoptimizationvariational}. Therefore, some variational computation approaches not guided strictly by an energy based cost function have been investigated for the task of high ground-state overlap state preparation. A core challenge for these approaches is that -- in order to guide a variational optimization process -- a cost function must be defined, and if ground-states (or, even low energy states) are not known, then there is no clear cost function to use. This is a significant issue because finding the ground-state is exactly the calculation that we are attempting to solve with these algorithms. One solution that has been proposed is to use overlap of higher-energy-states guided optimization~\cite{Feniou_2023, Feniou_2024}, however, this still requires some knowledge about the quantum Hamiltonian of interest in order to guide the optimization using intermediate overlap calculations. 

Thus, there are good reasons for skepticism that the VQE algorithm will produce high-overlap initial states for QPE~\cite{Louvet_2026, Fomichev_2024}. On the other hand, there are only a handful of quantum algorithms that can prepare good initial states, and VQE is one of the easiest to implement in terms of the quantum resources required. 
Our main hypothesis is that -- in light of the current growing community perspective that the VQE algorithm will not be able to reliably compute the ground-state of quantum Hamiltonians, unless either good scaling is identified or certain trainability issues are mitigated~\cite{Tilly_2022, Arrasmith_2021, Ragone_2024, Cerezo_2021, Anschuetz_2022, Wang_2021, McClean_2018, Larocca_2025, Uvarov_2021}, VQE could be used to efficiently prepare initial states with a reasonable overlap for subsequent use in QPE. VQE is known to be able to generate excited states of quantum Hamiltonians~\cite{Mizuta_2021, Smart_2024, grimsley2025challengingexcitedstatesadaptive, Benavides_Riveros_2024}.

Being used as an initial state preparation algorithm might be the only future large-scale quantum computational application for VQE. This therefore motivates the current study; we narrowly focus on a large set of extensive numerical calculations of running VQE training in order to assess how well it works for preparing high quality ground-states for QPE. We report extensive analysis of the limitations of using VQE for state preparation, different VQE ansatz choices, how entanglement (quantified by negativity) plays a role in what states VQE can prepare, and most importantly we empirically report the best-case scaling, obtained numerically (VQE run with shot noise), of the ground-state overlap. We demonstrate how VQE can fail to prepare good QPE initial states simply by ``autonomously'' finding low-energy states. The quantum Hamiltonian we consider is the $\pm J$ Heisenberg spin glass model (fully connected)~\cite{Itoi_2024},

\begin{equation}
\label{eq:heisenberg_spinglass}
\mathcal{H}
\,=\,
\sum_{i<j}
\Big(
  J^{x}_{ij}\,\sigma_i^{x}\sigma_j^{x}
 +J^{y}_{ij}\,\sigma_i^{y}\sigma_j^{y}
 +J^{z}_{ij}\,\sigma_i^{z}\sigma_j^{z}
\Big),
\end{equation}

where $J^{x}_{ij}, J^{y}_{ij}, J^{z}_{ij}$ weights are independently drawn uniformly at random to be either $+1$ or $-1$, making the model anisotropic. This Hamiltonian is a good, challenging, test case for VQE because finding ground-states of this model is hard; not only is it not geometrically local, the ground-states are entangled and can be degenerate, and there is a high degree of frustration. 

\begin{figure}[ht!]
    \centering
    \includegraphics[width=1.0\linewidth]{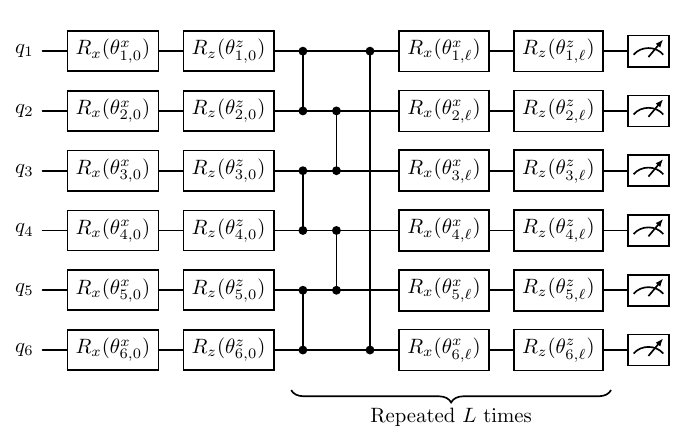}
    \caption{\textbf{VQE circuit diagram structure, applied to $6$ qubits.} $L$ parameterizes the VQE depth, where $L=0$ is the non-entangling basic case. The ansatz parameters are defined by the vector of $\theta_i \in (0, 2\pi)$ real values. In order to extract the full Hamiltonian expectation value, each set of parameters is evaluated using measurements in Pauli X, Y, and Z bases thus requiring basis change rotations prior to measurement. $\mathrm{CZ}$ gates are the entangling operation between pairs of qubits used in this ansatz.  }
    \label{fig:VQE_circuit}
\end{figure}

\section{Methods}
\label{section:methods}

\begin{figure*}[ht!]
    \centering
    \includegraphics[width=0.322\linewidth]{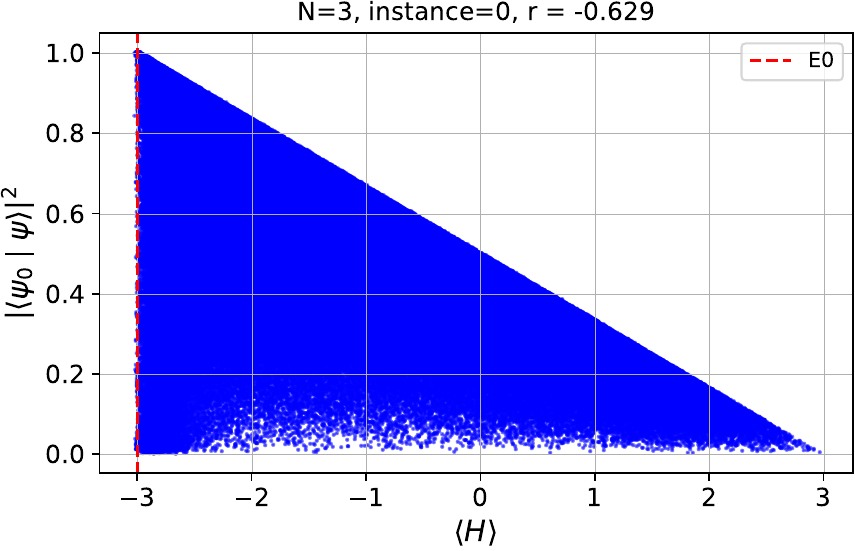}
    \includegraphics[width=0.322\linewidth]{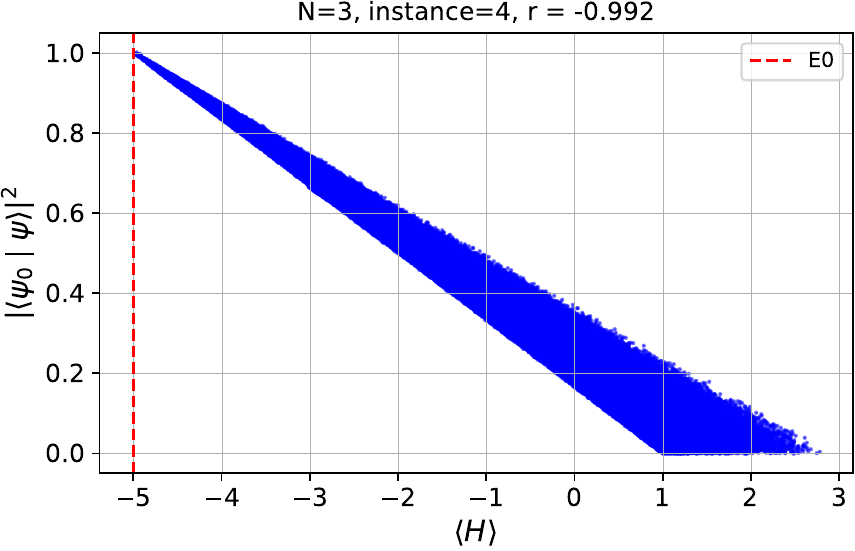}
    \includegraphics[width=0.322\linewidth]{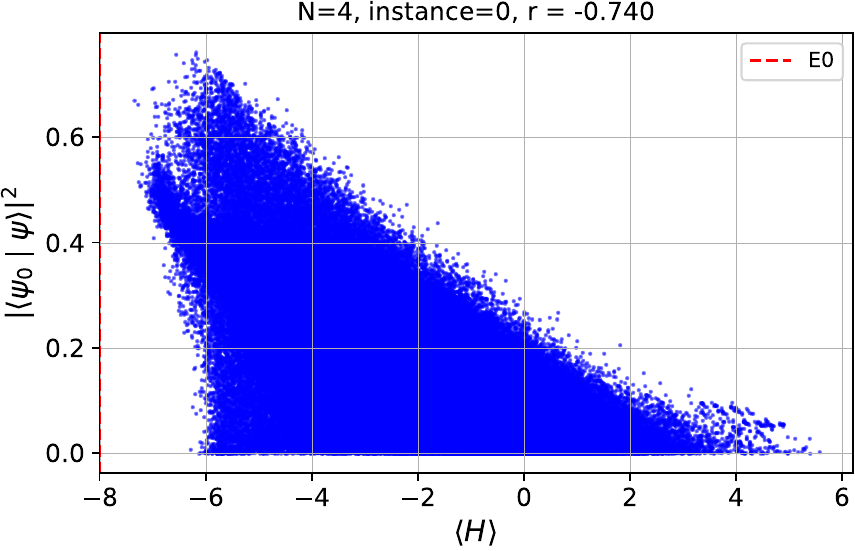}
    \includegraphics[width=0.322\linewidth]{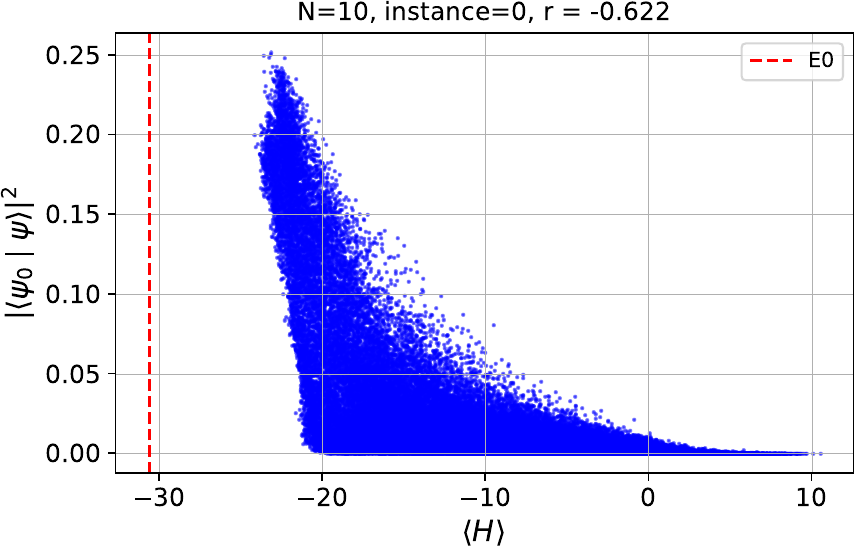}
    \includegraphics[width=0.322\linewidth]{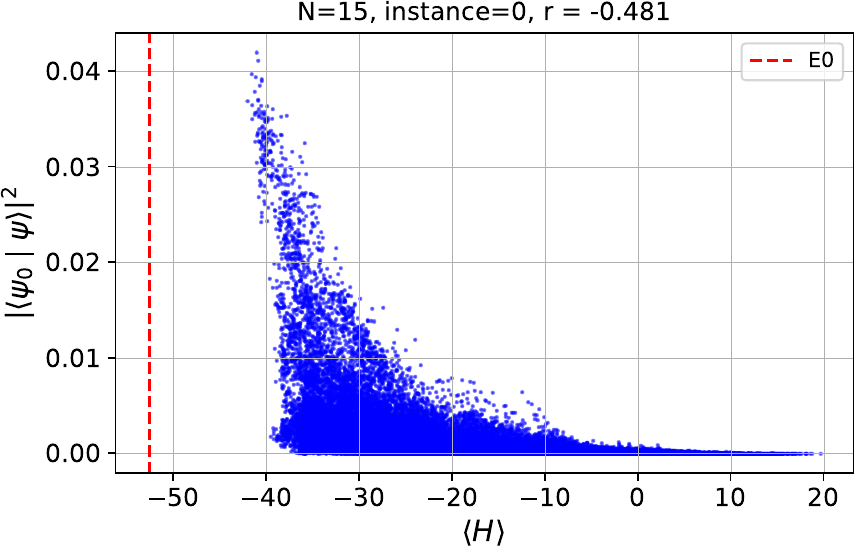}
    \includegraphics[width=0.322\linewidth]{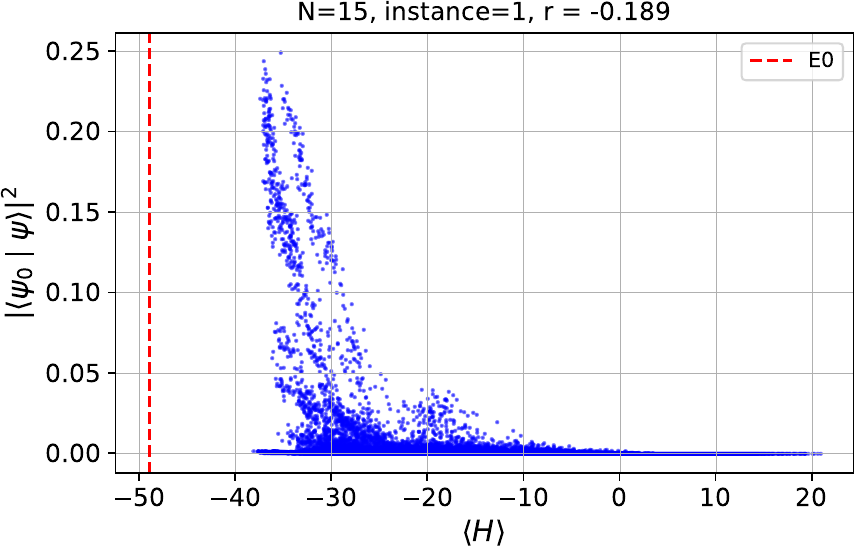}
    \caption{ \textbf{Energy (x-axis) vs overlap (y-axis) scatterplots of states produced during VQE training.} The 6 subplots each show a distribution produced from a different quantum Hamiltonian. The true minimum eigenvalue $E_0$ is denoted by a vertical dashed red line in each plot. The title of each subplot states the system size, an instance number index (which can range from $0$ to $9$), and the Pearson correlation coefficient ($r$) from a generic linear regression. The Hamiltonian expectation value $\langle H \rangle$ is obtained from the VQE training and therefore has shot noise, whereas $|\langle \psi_0 \mid \psi \rangle|^2$ is calculated exactly for each state, with no shot noise. ``Good'' states are in the upper left-hand corner of each plot (those with both higher overlap and low-energy).  }
    \label{fig:energy_vs_overlap}
\end{figure*}

The core goal of this study is to calculate ground-truth overlap scaling that VQE could produce in ideal settings. To this end, full exact diagonalization is required, for which we use the LAPACK linear algebra library~\cite{129995} within NumPy~\cite{harris2020array}. Degenerate ground-states are possible for this Heisenberg model. We use a floating point precision cutoff of $1 \times 10^{-15}$ in order to determine degeneracy. Degenerate ground states are all orthonormal eigenvectors as produced from the exact diagonalization routine. The overlap calculation must consider the degenerate eigenvectors, and therefore although the plot notation we will use is $| \langle \psi_0 \mid \psi \rangle|^2$ where $\psi_0$ is the ground-state and $\psi$ is the candidate (VQE-produced) state, the degeneracy-accounted overlap formula we use is $\sum_{\psi_0 \in v_0} \left| \langle \psi_0 \mid \psi \rangle \right|^2$ where $v_0$ is the set of orthonormal eigenvectors.

The VQE circuit ansatz is a standard type of ansatz where entangling layers are separated by parameterized single qubit rotations, in part based on the ansatz used in ref.~\cite{kirmani2025variationalquantumsimulationstwodimensional}. Two different entangling layers are tested - a 1D ring with periodic boundaries, and a clique ($\mathrm{CZ}$ gates between every pair of qubits in the complete graph). The $\mathrm{CZ}$ gate ordering is arbitrary within each entangling layer since they commute with each other. Fig.~\ref{fig:VQE_circuit} shows a rendering of the VQE circuit diagram using the 1D entangling layer and $6$ qubits.

We also test two different VQE parameter initialization distributions. The first is a uniform distribution from the interval $(0, 2\pi)$. The second is a Gaussian distribution with standard deviation of $\sqrt{1/(L+1)}$, centered at $0$~\cite{zhang2025escapingbarrenplateaugaussian}. There is evidence that variational algorithm initialization with the normal distribution whose standard deviation scales like $1/L$ has larger gradients compared to uniform parameter distribution~\cite{zhang2025escapingbarrenplateaugaussian}. The goal with testing both of these parameter distributions is to evaluate whether one is able to reach the ground-state, and therefore hopefully high overlap states, better than the other. Becoming stuck in local minima when performing the optimization routine of VQE is very common~\cite{kirmani2025variationalquantumsimulationstwodimensional, atallah2025investigatingdifferentbarrenplateaus, Sannia_2024, belaloui2024groundstateenergyestimation}, and for this reason in these tests, for each unique quantum Hamiltonian we use $20$ different random parameter initializations, VQE depth $L=0, \ldots, 20$, both clique and ring entangling layers, and lastly both uniform and Gaussian parameter initialization. The $L=0$ case is intended to test the degree to which entanglement helps the training, and provides a classical probability distribution baseline. This extensive numerical testing is performed with the goal of thoroughly evaluating whether VQE can perform well under different ansatz and parameter distribution conditions at this state preparation task. The largest VQE circuit, defined on $15$ qubits, that we utilize is composed of $2100$ $\mathrm{CZ}$ gates and $630$ independently trainable single qubit rotation gates ($\mathrm{R}_z$ and $\mathrm{R}_x$ gates).

The classical optimizer used to learn the VQE parameters is Simultaneous Perturbation Stochastic Approximation (SPSA)~\cite{657661, spall2002multivariate, Kandala_2017}, with $100$ learning steps, corresponding to exactly $251$ objective function evaluations, and all other hyperparameters are left to default. SPSA is a good general purpose black-box optimizer, useful for cases where the underlying gradient is not known, that is able to handle stochasticity due to shot noise. The Hamiltonian expectation value produced by a single VQE parameter combination is evaluated using $100{,}000$ shots (measurements) for each of the three Pauli bases (X, Y, Z). These circuit executions all use the statevector Qiskit~\cite{javadiabhari2024quantumcomputingqiskit} simulator with no error model (the only source of error is finite sampling). The inclusion of shot noise in the training is intended to make the results more realistic, since any real quantum computer training will necessarily include substantial shot noise. The objective function used in training is strictly a minimization of the energy of the Hamiltonian. In other words, this is very intentionally a standard VQE setup. These numerical calculations required substantial HPC (High Performance Computing) compute to be run concurrently. 

In post-processing, the overlap of each state that VQE prepares is calculated numerically with no shot noise, as is the entanglement measure negativity. Negativity~\cite{PhysRevA.58.883, Vidal_2002} is defined as

\begin{equation}
\mathcal{N}(\rho) = \frac{\vert \vert \rho^{T_A} \vert \vert_1 - 1 } {2},
\end{equation}

where $T_A$ is the partial transpose on subsystem $A$ of the density matrix $\rho$. We use a subsystem $A$ that is an arbitrarily chosen half of the qubits of the Hamiltonian $\lfloor{N/2} \rfloor$. The negativity on a density matrix $\rho$ is calculated using the Python package Toqito~\cite{toqito}, and the density matrix is calculated exactly with no shot noise or state tomography in Qiskit~\cite{javadiabhari2024quantumcomputingqiskit}. When plotting the negativity we normalize it by dividing by the maximum possible negativity, which is $\frac{2^{\lfloor N/2 \rfloor} - 1}{2}$.

\begin{figure*}[ht!]
    \centering
    \includegraphics[width=0.49\linewidth]{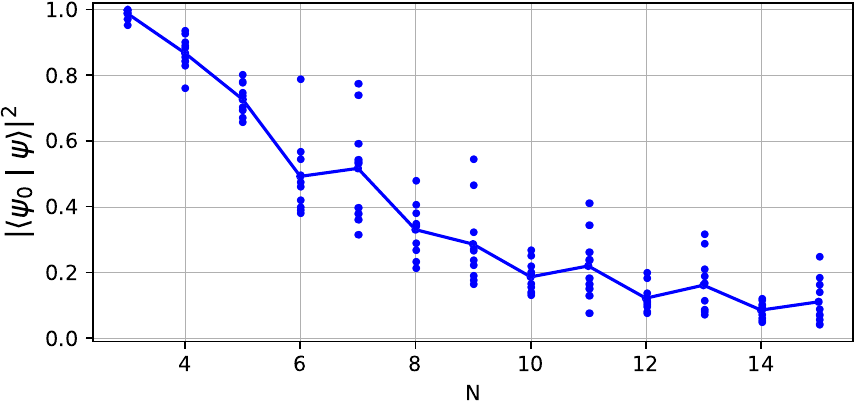}
    \includegraphics[width=0.49\linewidth]{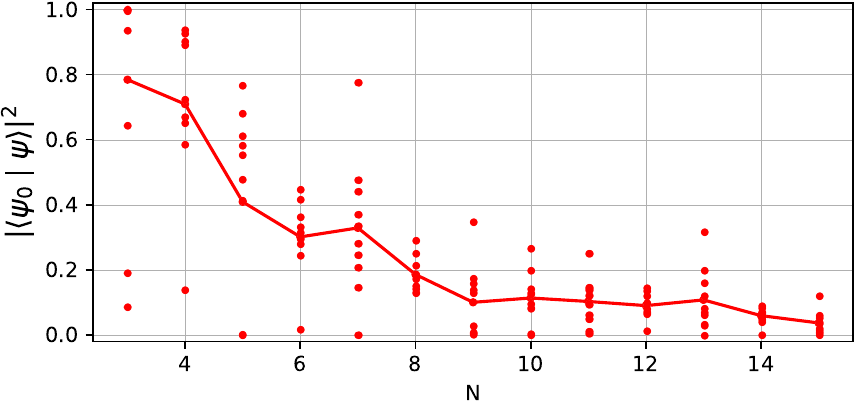}
    \includegraphics[width=0.49\linewidth]{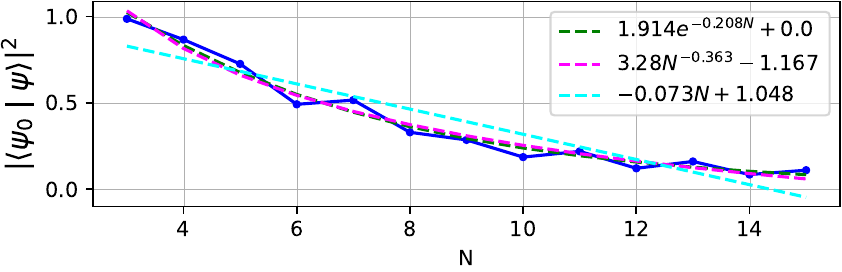}
    \includegraphics[width=0.49\linewidth]{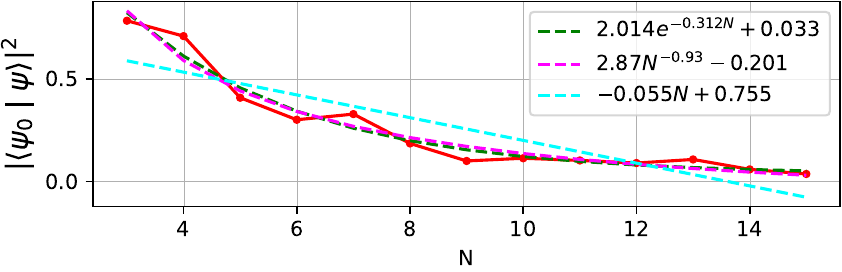}
    \caption{ \textbf{Top left: Largest overlap computed from VQE as a function of quantum Hamiltonian size (x-axis), top right: overlap of the absolute lowest energy VQE state as a function of quantum Hamiltonian size.} For each system size there are $10$ independent instances with different coefficients, here each point in the left subplot is the absolute best overlap found during VQE training for each instance. The solid line plots the mean overlap among the $10$ instances. The energy estimate used to determine the lowest energy VQE state for the right-most plot is calculated using the finite-shot estimate from the training. Bottom row plots best-fit regression lines for exponential (3 parameters), polynomial (3 parameters), and linear (2 parameters), fits to the average overlap.  }
    \label{fig:best_overlap_scaling}
\end{figure*}

\begin{figure}[ht!]
    \centering
    \includegraphics[width=1.0\linewidth]{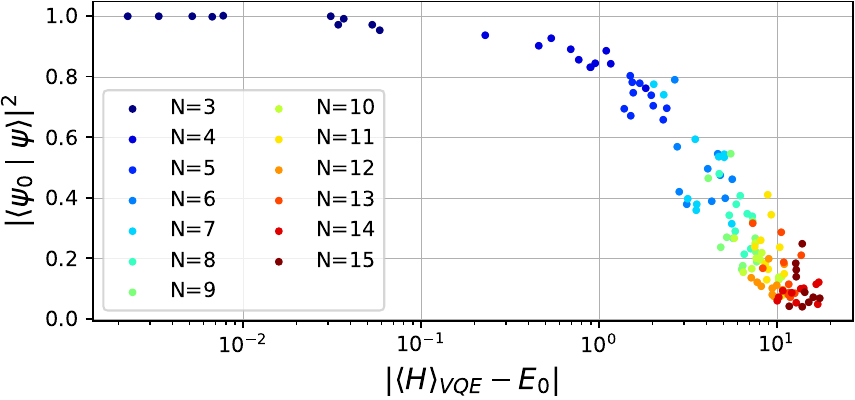}
    \caption{ \textbf{Highest overlap VQE computed state as a function of the energy absolute-error of that state (log scale x-axis).} Each plotted dot corresponds to a single quantum Hamiltonian. The estimate of $\langle H \rangle_{VQE}$ has shot noise, whereas both $E_0$ and $|\langle \psi_0 \mid \psi \rangle|^2$ are calculated exactly with no shot noise. }
    \label{fig:energy_error_vs_overlap}
\end{figure}

\begin{figure*}[ht!]
    \centering
    \includegraphics[width=0.49\linewidth]{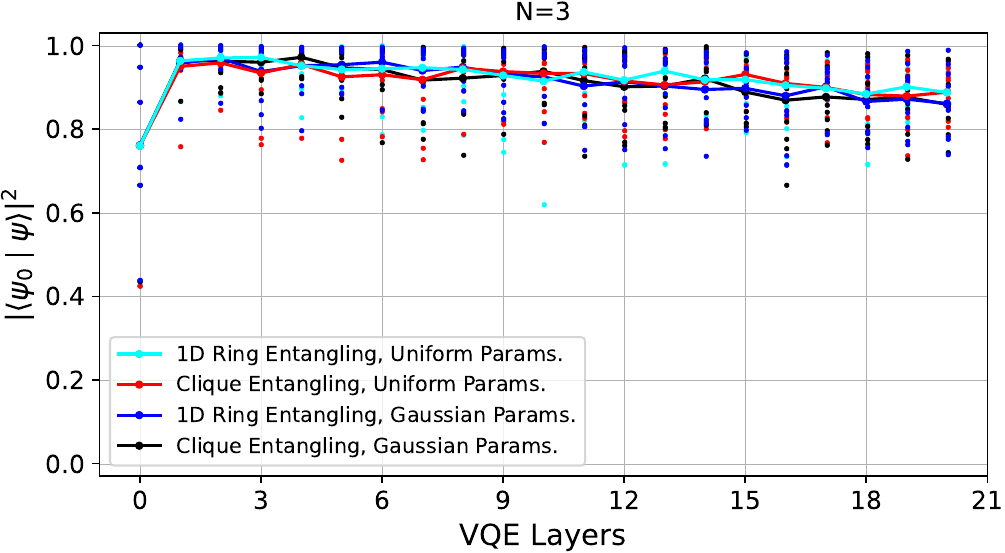}
    \includegraphics[width=0.49\linewidth]{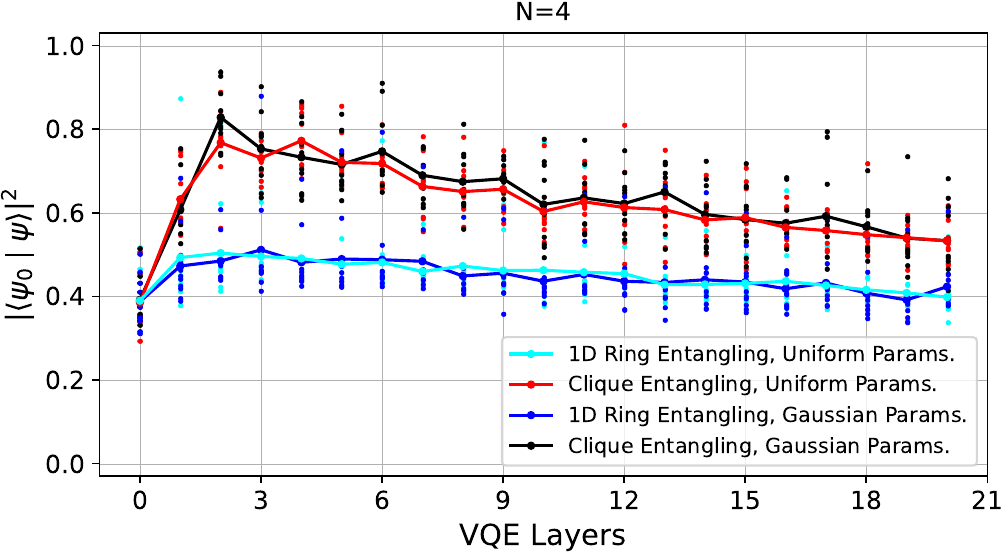}
    \includegraphics[width=0.49\linewidth]{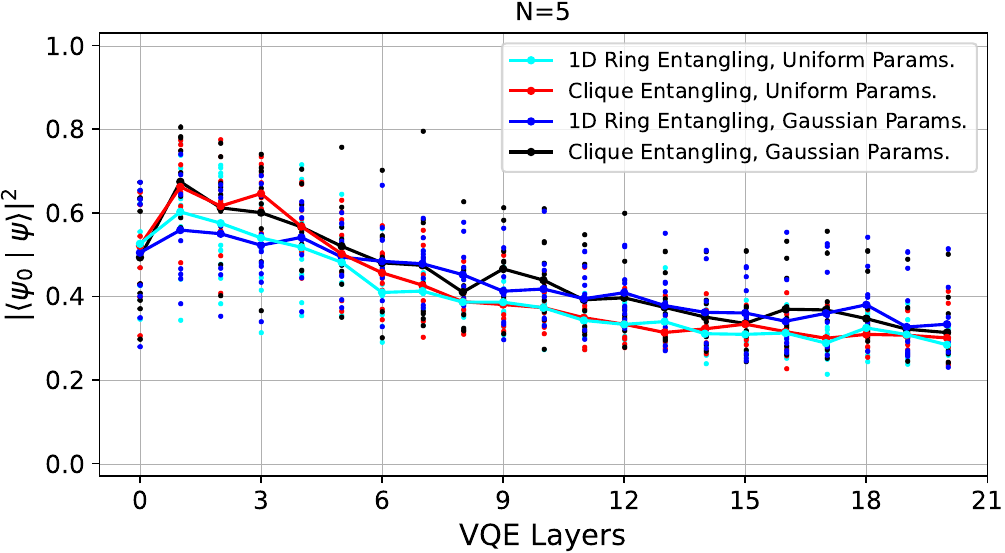}
    \includegraphics[width=0.49\linewidth]{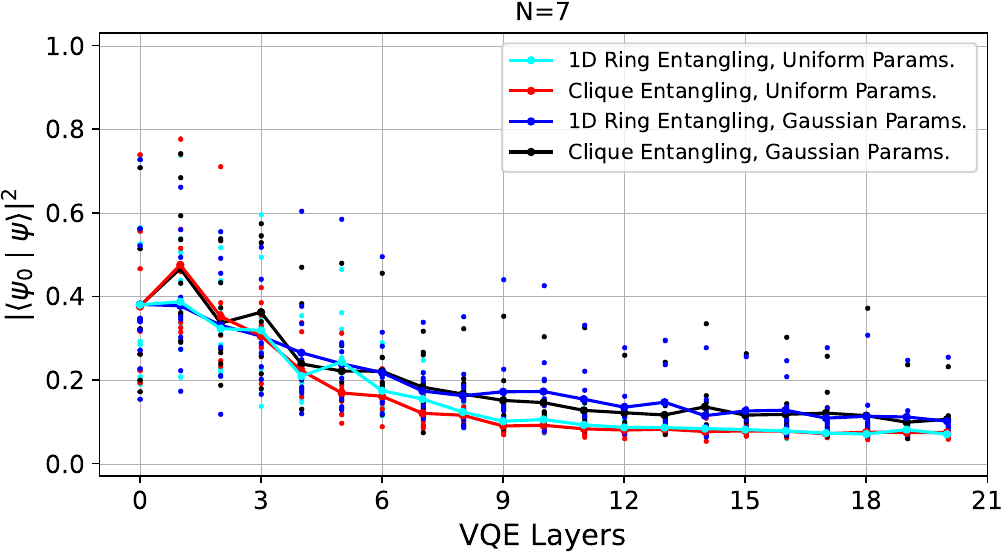}
    \includegraphics[width=0.49\linewidth]{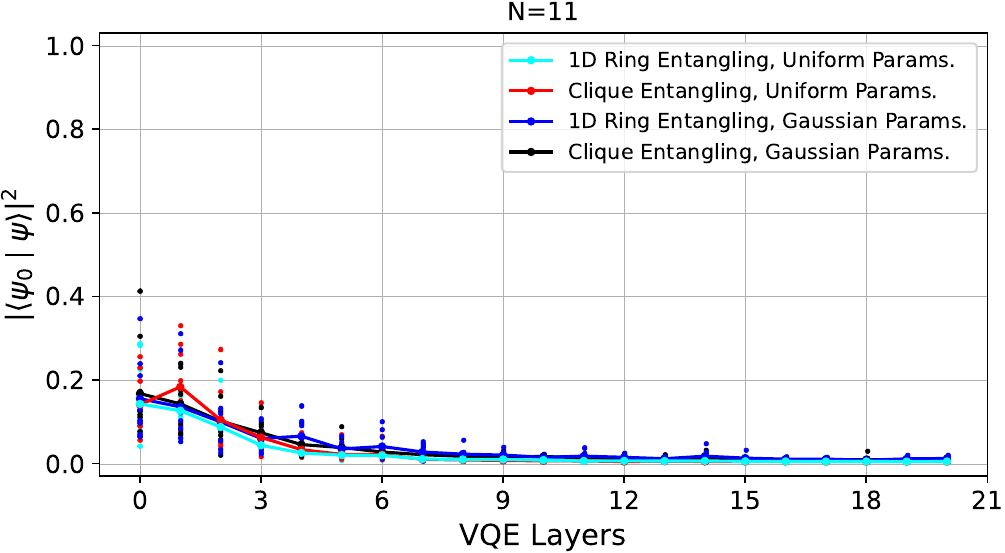}
    \includegraphics[width=0.49\linewidth]{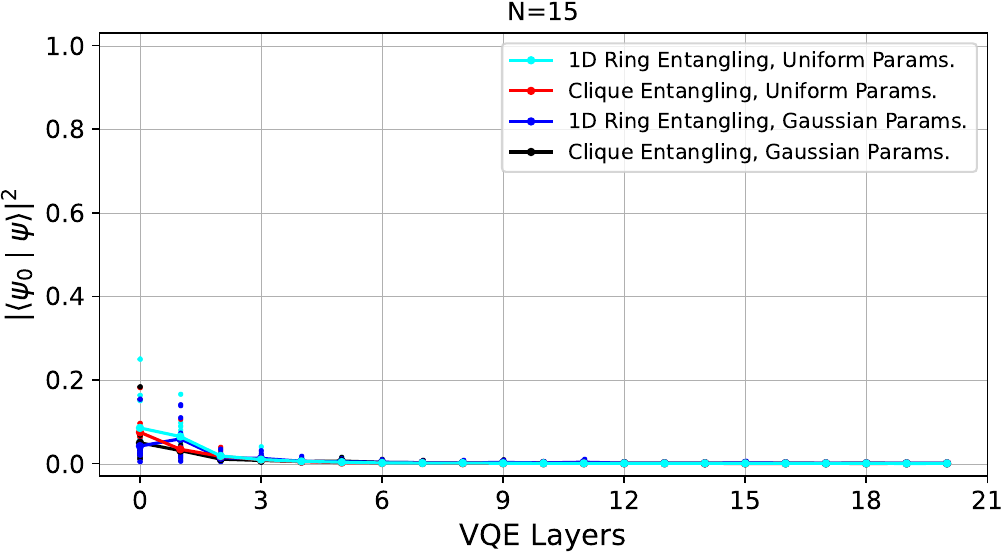}
    \caption{ \textbf{Comparing the performance of number of entangling layers, ring and clique entangling layers, and parameter initialization distributions with respect to overlap.} This is an absolute-idealized overlap case, where only the best overlap found from each numerical computation (for each ansatz) is plotted.  }
    \label{fig:function_of_layers}
\end{figure*}

\begin{figure*}[ht!]
    \centering
    \includegraphics[width=0.49\linewidth]{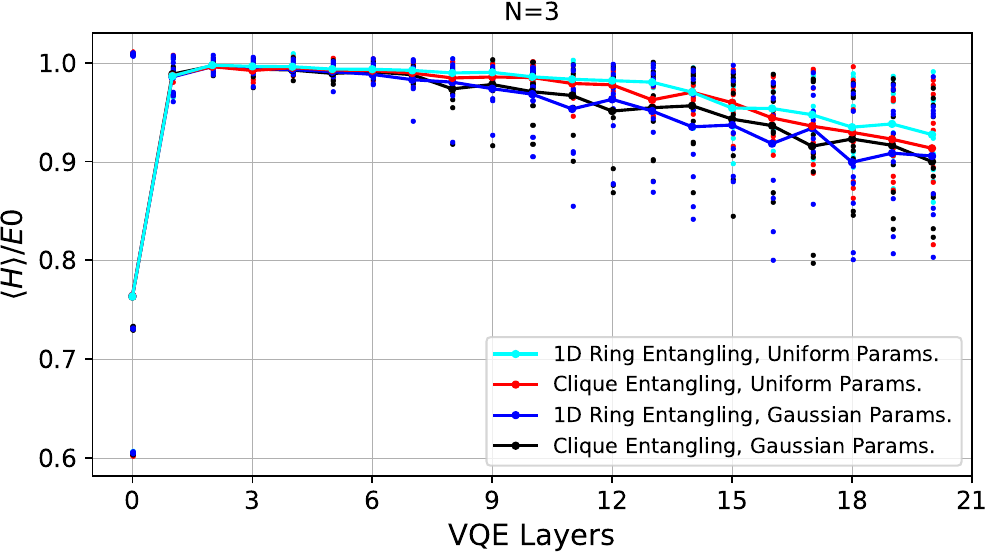}
    \includegraphics[width=0.49\linewidth]{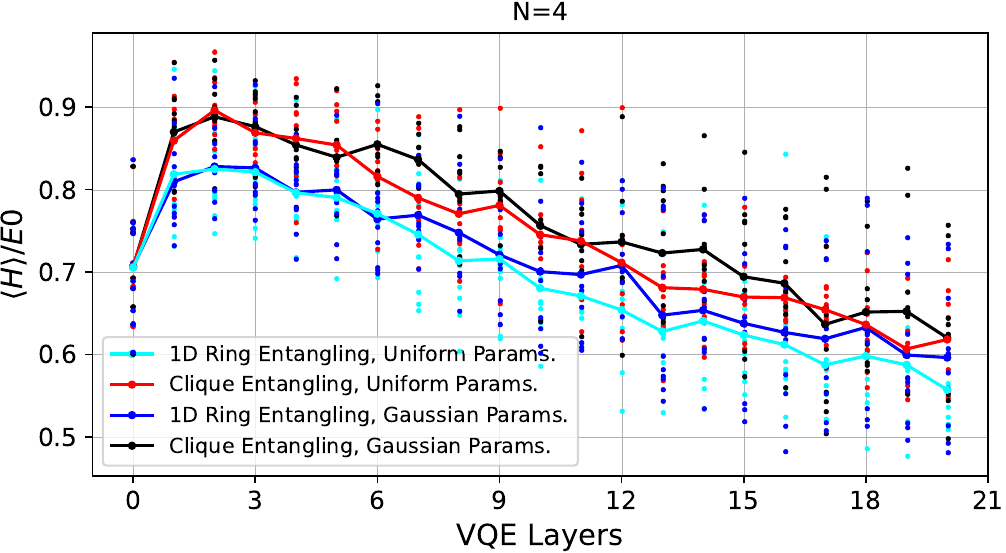}
    \includegraphics[width=0.49\linewidth]{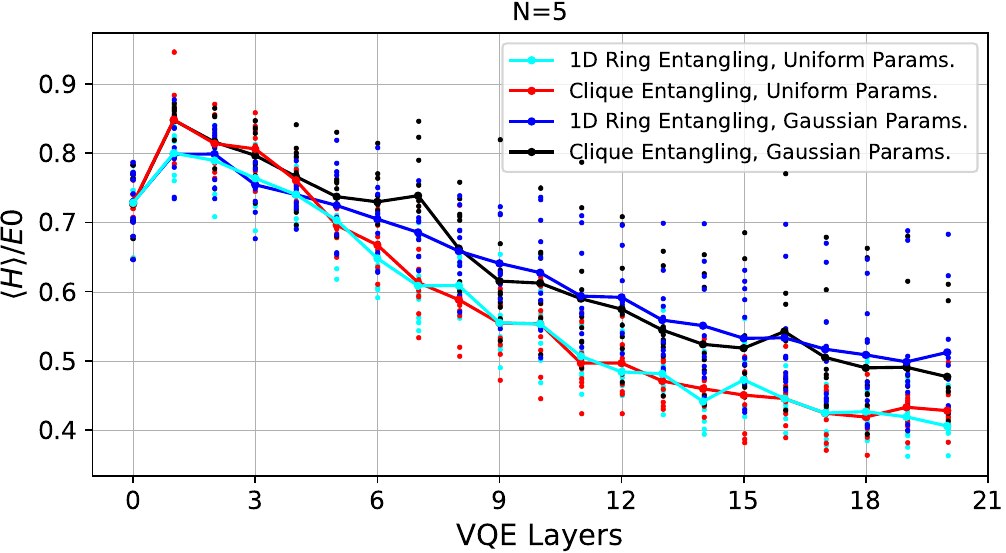}
    \includegraphics[width=0.49\linewidth]{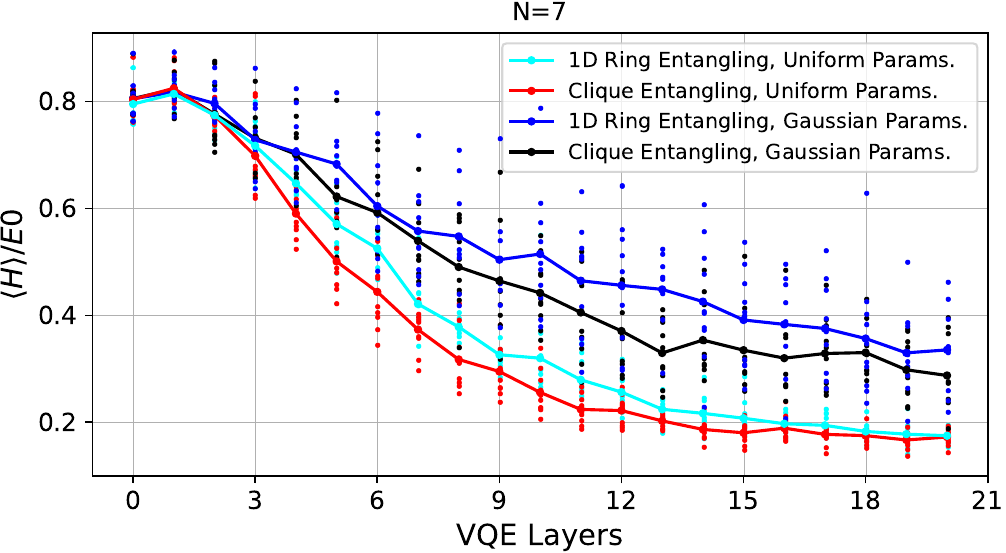}
    \includegraphics[width=0.49\linewidth]{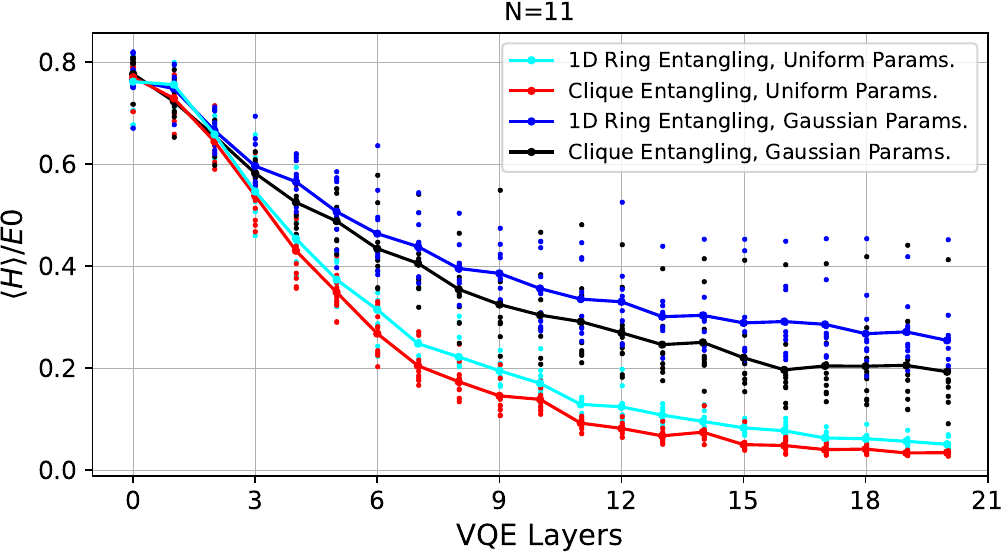}
    \includegraphics[width=0.49\linewidth]{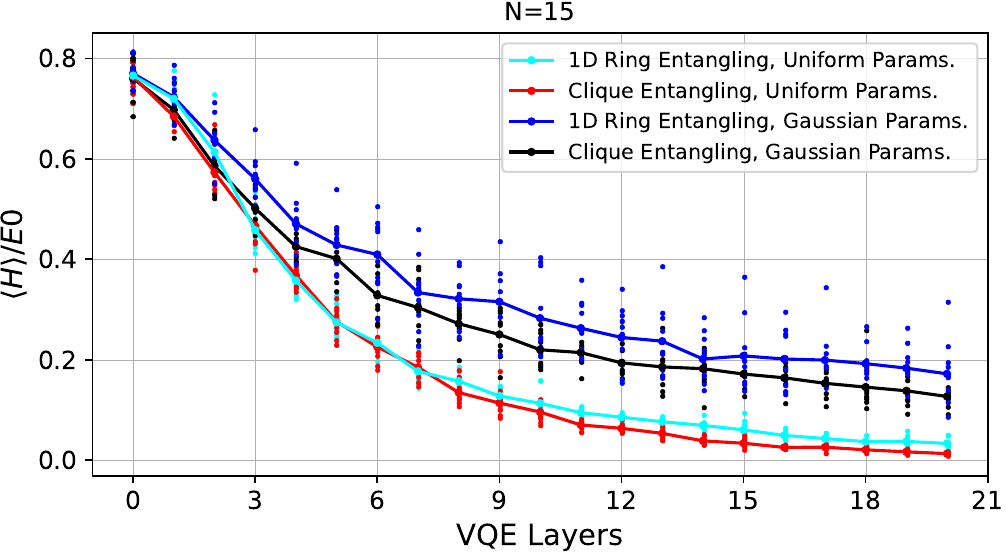}
    \caption{ \textbf{VQE performance, with varying ansatz and number of layers, with respect to normalized energy.} The best possible value on the y-axis is $1.0$ (which corresponds to the ground-state energy). Interestingly, this shows that at higher ansatz depths the trainability fails substantially, so much so that at larger $N$ the most successful ansatz is typically $L=0$ (no entanglement) or $L=1$.  
    }
    \label{fig:function_of_layers_energy}
\end{figure*}

\section{Results}
\label{section:results}

The first question to address is how the energy cost function correlates with overlap. Fig.~\ref{fig:energy_vs_overlap} plots exactly this relationship for several representative quantum Hamiltonian instances of varying sizes. This shows that there is a linear correlation between Hamiltonian energy and overlap - in some cases this correlation is strongly linear with a correlation coefficient magnitude of $0.992$, whereas in other cases the correlation is only weakly linear ($<0.5$). Importantly this clearly illustrates that in some cases, although the energy can be very near to the true ground-state, the overlap may be zero or close to zero. Each subplot of Fig.~\ref{fig:energy_vs_overlap} reports the full distribution of evaluated expectation values during VQE training, including all combinations of entangling layers and parameter initializations -- each distribution is composed of exactly $21 \cdot 2 \cdot 2 \cdot 251 \cdot 20 = 421{,}680$ unique states ($20$ random parameter starts, $0$ to $20$ entangling layers, $251$ objective function calls during training, $2$ different entangling layer types, and $2$ different parameter distributions). The problem is that clearly energy does not always correlate to ground-state overlap, which means that when operating VQE, using strictly the minimum energy found does not guarantee high overlap. Moreover, there is no efficient way to test the state for how high the overlap is.

The other key thing that is clear from Fig.~\ref{fig:energy_vs_overlap} is that there is a strong dependence on system size. Fig.~\ref{fig:best_overlap_scaling}-(left) examines this in more detail by plotting the absolute best overlap computed during VQE training as a function of system size. This shows that there is a dramatic dropoff of overlap as a function of $N$. However, this decrease is not clearly exponential. As a point of comparison, ref.~\cite{pelofske2026numericalexperimentsparametersetting} reported overlap scaling as a function of $N$ for a handful of easy-to-prepare states such as GHZ states, graph states, and staggered X gate states - the result was that the overlap scaling was clearly strongly exponentially decreasing, with all of those states having an overlap of approximately $\approx 10^{-4}$ at $N=14$. Thus, the VQE prepared states clearly have a substantially better overlap than the simple states considered in ref.~\cite{pelofske2026numericalexperimentsparametersetting} for fully-connected Heisenberg spin-glass models. Larger system sizes, much larger than $N=15$, would need to be rigorously evaluated in order to strongly determine what the asymptotic scaling is -- at this sample size the true scaling is not possible to determine with confidence. Fig.~\ref{fig:best_overlap_scaling} confirms this by showing that whether the scaling is exponentially decreasing or is a simple polynomial, is ambiguous (although, it is clearly not a simple linear scaling). The root mean squared error (RMSE) of the polynomial fits are $0.04722$ (Fig.~\ref{fig:best_overlap_scaling}-left) and $0.04916$ (Fig.~\ref{fig:best_overlap_scaling}-right) and the exponential fits are $0.04044$ (Fig.~\ref{fig:best_overlap_scaling}-left) and $0.04379$ (Fig.~\ref{fig:best_overlap_scaling}-right). Determining the true asymptotic scaling numerically is also challenging for VQE because ultimately it is a form of machine learning, and it is unclear what, if any, large-scale trainability barriers exist.

\begin{figure*}[ht!]
    \centering
    \includegraphics[width=0.32\linewidth]{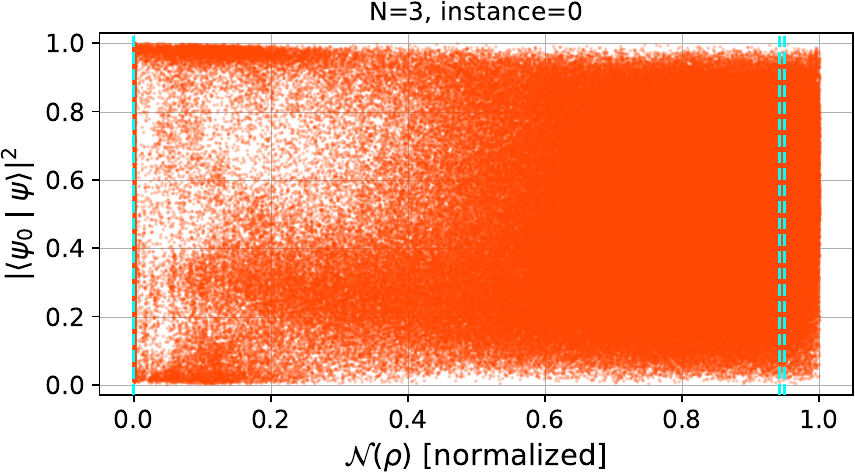}
    \includegraphics[width=0.32\linewidth]{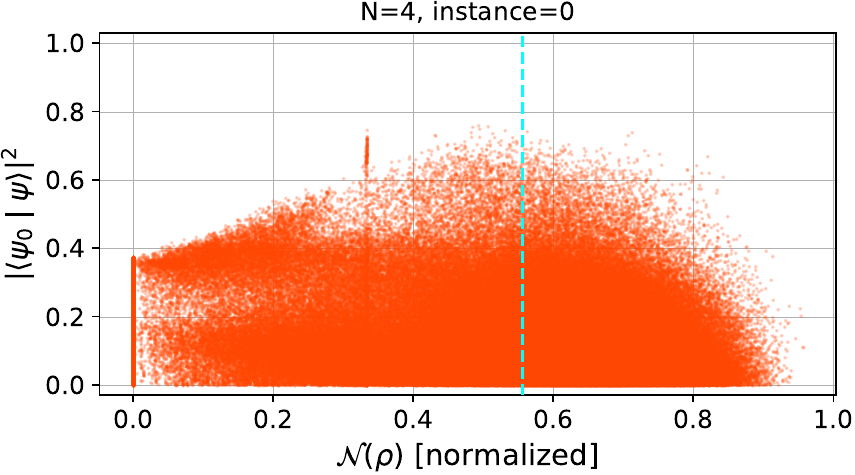}
    \includegraphics[width=0.32\linewidth]{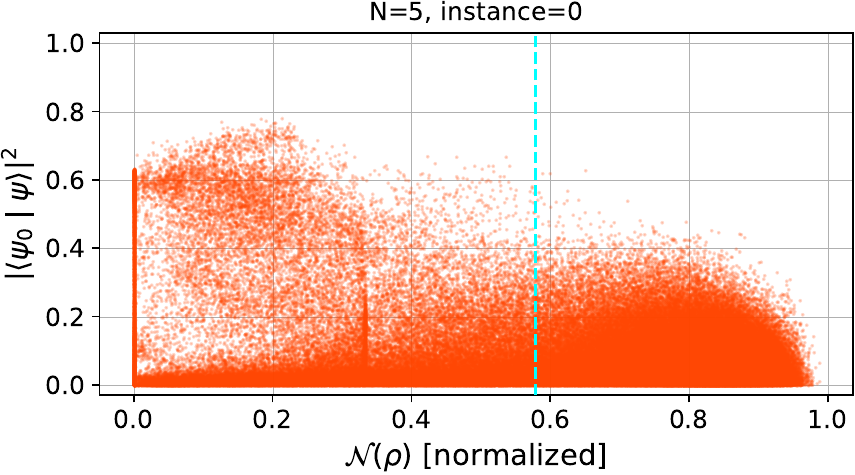}
    \includegraphics[width=0.32\linewidth]{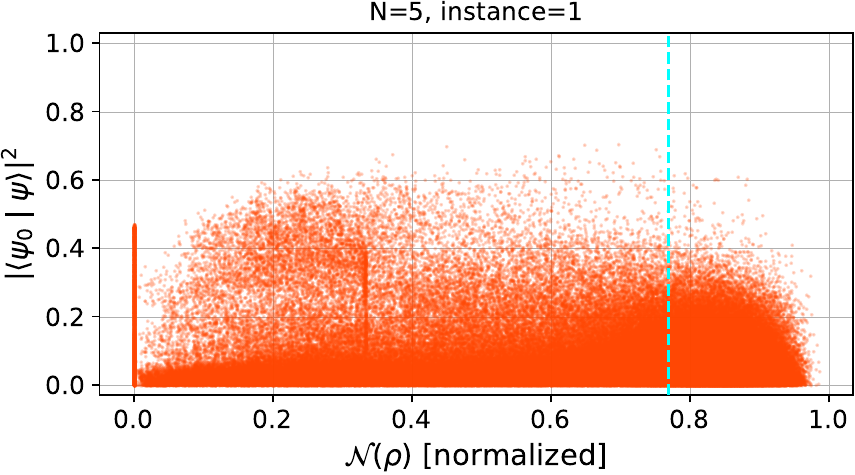}
    \includegraphics[width=0.32\linewidth]{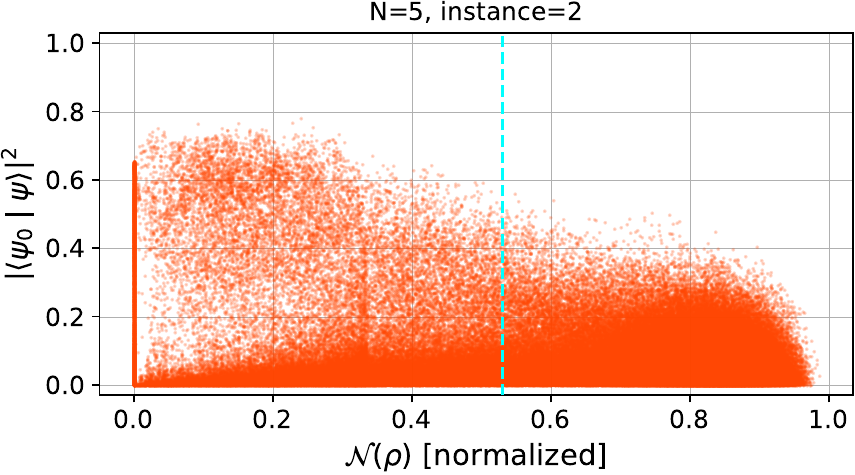}
    \includegraphics[width=0.32\linewidth]{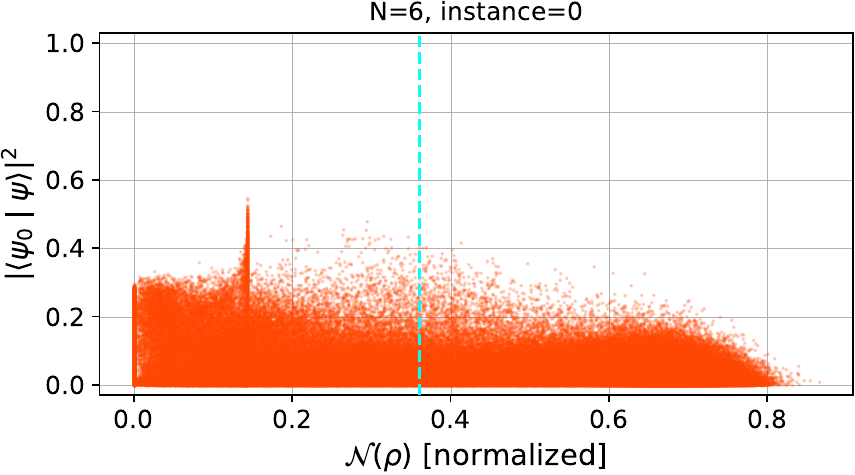}
    \includegraphics[width=0.32\linewidth]{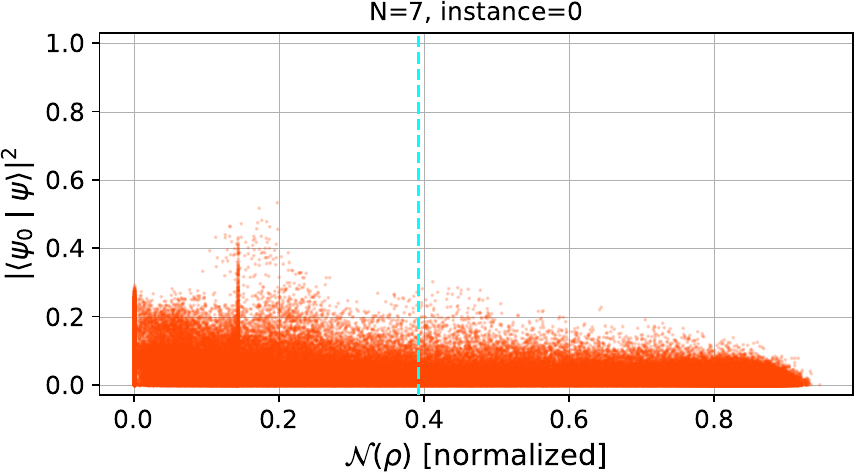}
    \includegraphics[width=0.32\linewidth]{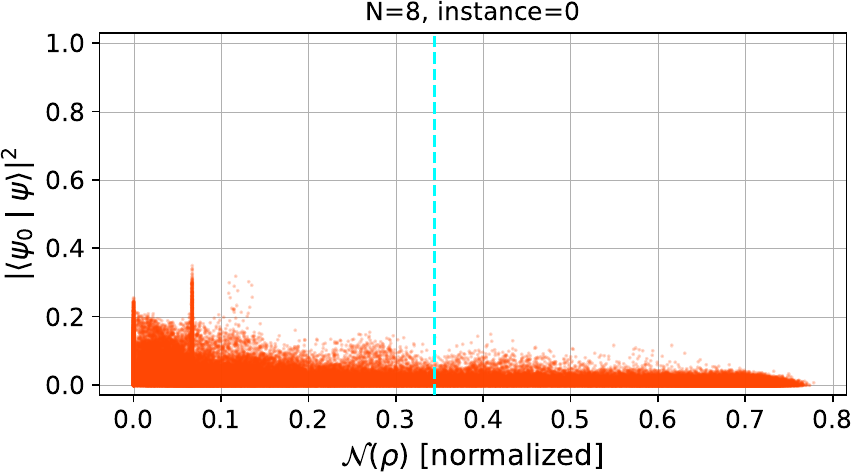}
    \includegraphics[width=0.32\linewidth]{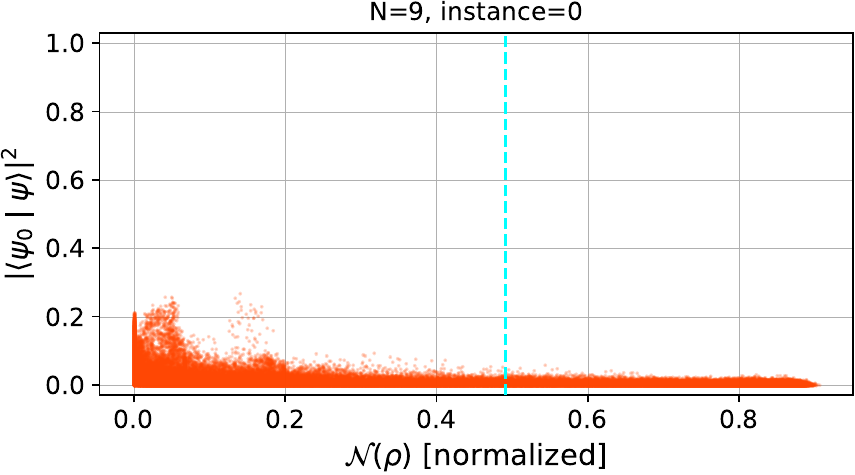}
    \caption{ \textbf{VQE produced state entanglement negativity (x-axis) and overlap (y-axis) distributions.} Each orange point corresponds to a single VQE produced state, and the point-cloud in each distribution defines the complete set of VQE produced states during training for each quantum Hamiltonian. Vertical cyan dashed lines denote the negativity of the true ground-state (multiple lines denote degenerate ground-states), from the particular orthonormal eigenvectors produced by the exact diagonalization. The maximum possible negativity is $1.0$ (since it is normalized). Neither plotted quantity has shot noise.  }
    \label{fig:negativity}
\end{figure*}

Fig.~\ref{fig:best_overlap_scaling}-(left) examines an \emph{idealized} scaling of the best overlap found from VQE. However, given that there is no way to determine which state has the high overlap, Fig.~\ref{fig:best_overlap_scaling}-(right) plots the overlap of the absolute lowest-energy found VQE state as a function of $N$, in which case we see a clear decrease in overlap in particular for some Hamiltonians the overlap is very near to zero. This shows that purely relying on the Hamiltonian expectation value is not sufficient in general to guide towards a high-overlap state. One possible work-around is to generate an ensemble of VQE produced states with energy $\langle H \rangle < 0$ (at least, in the case of this $\pm J$ Heisenberg model, see Fig.~\ref{fig:energy_vs_overlap}), and then try each of those states individually as initial states of QPE -- as long as one of those VQE states has sufficient ground-state overlap, the minimum eigenvalue of the quantum Hamiltonian will be able to be computed by QPE.

Still, the overlap approximately converging to roughly $0.1$ is tentative positive numerical evidence that VQE could produce states with reasonably good overlap at larger problem sizes. Fig.~\ref{fig:energy_error_vs_overlap} plots the absolute highest overlap state as a function of the absolute error, in energy, of that state as compared to the true ground-state energy $E_0$. The key thing that this plot shows is that the absolute error in energy grows quite substantially as a function of $N$ -- for the largest quantum Hamiltonians the absolute error is $\approx 10$. Despite these high error rates, the (best) overlap remains relatively high.

Fig.~\ref{fig:function_of_layers} plots the best-overlap found as a function of the number of entangling layers in the VQE ansatz. This shows that there is no universal winning parameter combination, although typically the clique entangling structure works better than the ring entangling structure, which is physically consistent with the clique structure of the quantum Hamiltonian. Interestingly, the Gaussian parameter initialization does not clearly perform better than the uniform parameter initialization, which provides some numerical evidence against gradient vanishing being the primary trainability issue for VQE. The higher depth VQE circuits fail to find low-energy states of the quantum Hamiltonian. For instance it is clear that $L=1, 2, 3$ performs much better than $L=20$. Higher VQE depths failing to perform well is a notable result of these numerical simulations because they were performed in very idealized settings (e.g., although shot noise is present, there is no other source of error or noise). For a point of comparison, Fig.~\ref{fig:function_of_layers_energy} plots strictly the minimum-energy performance of VQE, which shows that a similar trend occurs as in the case of the overlap in Fig.~\ref{fig:function_of_layers}. This demonstrates that the VQE training is failing, likely due to a combination of the underlying energy landscape of the VQE parameters being hard to search over and also could be due to the ground-state of this quantum Hamiltonian being inherently hard to approximate due to non-local connections and frustration. This means there could exist a classical optimization routine with better success than SPSA. Still, VQE failing at higher ansatz depths is a relatively rare observation (Figure 2 of ref~\cite{stober2021considerationsevaluatingthermodynamicproperties} is one example), with most VQE typically being shown to continue to improve at higher depths~\cite{Tkachenko_2021, Long_2024, wu2025enhancingreachabilityvariationalquantum}. 

Interestingly, Fig.~\ref{fig:function_of_layers_energy} shows that at larger $N$ and high depth, the best performing parameter combination is the 1D entangling structure and Gaussian parameter initialization. This is notable because the clique entangling structure matches the connectivity of the Hamiltonian, however despite this the training is clearly harder in the clique entangling ansatz case. This heavily suggests that the non-local correlations that arise from the denser entangling layers cause the classical optimization to fail more than a simpler entangling structure.

Higher-depth VQE ansatzes have substantially more expressibility, however, they are also much harder to train. Moreover, at high depths, the ansatz can become over-parameterized, leading to a different type of barren plateau (one not induced by noise or decoherence)~\cite{hashimoto2026comprehensivenumericalstudiesbarren, you2022convergencetheoryoverparameterizedvariational}. In this case, although we do not systematically evaluate this, it is very likely that the ansatz at $L=20$ is substantially over-parameterized. Therefore this failure mode is due to the optimizer failing, not due to a lack of expressibility. Moreover, this is a well known phenomenon, where VQE circuits with a large number of parameters become very hard to train. Something to note is that the overlap obtained by $L=0$ is non-negligible in most cases, and performs better than the over-parameterized $L>10$ ansatz when $N \geq 5$. This is notable because this state is entirely classical, a product state, with no entanglement, and we expect that these product states will have exponentially decreasing ground-state overlap as system sizes increase. Moreover, the $L=0$ case gives a baseline showing the benefit of having entanglement in the VQE ansatz -- this benefit of having entanglement decreases as $N$ increases, demonstrating that the VQE training struggles substantially as $N$ and $L$ both increase. This necessitates developing either easier to train VQE ansatzes or better classical optimization routines if VQE is ever going to be used for large scale state preparation, because $L=0$ VQE does not use any entanglement and therefore does not need to be trained on a quantum computer (this could be trained on a purely classical computer as a simple probability distribution modeling each qubit). 

One last perspective we examine is that of entanglement. Fig.~\ref{fig:negativity} reports the bipartite negativity distribution of the VQE produced states as a function of the state overlap, and with the negativity of the ground-state(s) marked. This shows that VQE is able to produce states that are highly entangled, and also states with entanglement very close to the entanglement of the ground-state. However, the entanglement is very uncorrelated with the overlap, even if the entangling structure of the VQE ansatz shares the same structure as the Hamiltonian (in the case of the clique entangling graph). This shows that merely the ability to prepare states with similar entanglement as the ground-state is not sufficient to produce states with high ground-state overlap.

\section{Discussion and Conclusion}
\label{section:conclusion}

We have numerically shown that for up to $N=15$ qubit Heisenberg model quantum Hamiltonians, VQE can prepare states with reasonably high ground-state overlap. Importantly, in order for this overlap property to be of use for QPE, it is not required that VQE necessarily get very close to the true ground-state, which is good because typically VQE fails to approximate ground-states very well due to a variety of trainability issues. The true scaling of the overlap up to larger system sizes (e.g., hundreds of qubits) is hard to determine from this small system size numerical study, but the scaling appears potentially favorable (specifically, non-exponentially decreasing). This suggests that the primary useful application of VQE could be as initial state preparation for QPE in the ``early fault-tolerant era''. Moreover, beyond ``vanilla VQE'', there are techniques such as \emph{state preparation boosters}~\cite{Wang_2022} which could improve on the overlap of a VQE produced state. 

An important practical consideration of using VQE for state preparation is that there is currently no clear way to exactly verify the overlap of the VQE prepared state; the VQE state would be directly used as the prepared state for the time evolution register of the QPE algorithm. This means that not only can no overlap property be verified (because that would require knowing the ground-state, which is the very problem we wish to solve), but also the full VQE state can not be efficiently characterized (e.g., quantum state tomography). This means that in practice, a lowest-energy VQE state would need to be used as an initial state, with no verification. It may be advisable to utilize a number of different low-energy VQE states as QPE initial states, in independent QPE executions, since there is not necessarily a direct correlation between low-energies and high-overlap. Future work should consider similar numerical algorithmic studies to quantify how well other similar types of variational and relatively short-depth quantum algorithms could feasibly function for initial state preparation with high ground-state overlap.

\section*{Acknowledgments}
\label{sec:acknowledgments}
The authors thank Yigit Subasi for many productive discussions and Marco Cerezo for helpful feedback on vanishing gradients. This work was supported by the U.S. Department of Energy through the Los Alamos National Laboratory. Los Alamos National Laboratory is operated by Triad National Security, LLC, for the National Nuclear Security Administration of the U.S. Department of Energy (Contract No. 89233218CNA000001). This research used resources provided by the Los Alamos National Laboratory Institutional Computing Program. This research used resources provided by the Darwin testbed at Los Alamos National Laboratory (LANL) which is funded by the Computational Systems and Software Environments subprogram of LANL's Advanced Simulation and Computing program (NNSA/DOE). Research presented in this article was supported by the NNSA's Advanced Simulation and Computing Beyond Moore's Law Program at Los Alamos National Laboratory and by the Laboratory Directed Research and Development program of Los Alamos National Laboratory under project number 20240083DR. LA-UR-26-24852.

\bibliographystyle{apsrev4-2-titles}
\bibliography{references}
\end{document}